# A Blueprint for Building Serverless Applications on the Net

A. I. Khan, R. Spindler

School of Network Computing
Monash University
Frankston, Australia
`Asad.Khan@infotech.monash.edu.au`

## Abstract

*A peer-to-peer application architecture is proposed that has the potential to eliminate the back-end servers for hosting services on the Internet. The proposed application architecture has been modeled as a distributed system for delivering an Internet service. The service thus created, though chaotic and fraught with uncertainties, would be highly scalable and capable of achieving unprecedented levels of robustness and viability with the increase in the number of the users. The core issues relating to the architecture, such as service discovery, distributed application architecture components, and inter-application communications, have been analysed. It is shown that the communications for the coordination of various functions, among the cooperating instances of the application, may be optimised using a divide-and-conquer strategy. Finally, the areas where future work needs to be directed have been identified.*

## 1. Introduction

The Internet, with its roots in the ARPANET, was primarily conceived as a means of remote login and experimentation with data communication protocols. Hence there is invariably a challenge when it comes to sustaining services on the basis of strict financial rationalisation. Generally the services that are deemed as successful are the ones with a large number of users base and are provisioned free of cost to the users. Some of these freeware have the potential of generating revenue through commercial advertising however not all applications can satisfactorily sustain themselves through advertisement based income alone. As a consequence it becomes rather difficult to sustain such applications in spite of these services being of considerable community interest. After a while the market forces either puts the application out of service or the few isolated servers become overly subscribed rendering the service unusable. Hence comes the desire to explore possibilities where such applications may be made economically viable by doing away with the demands for more server-based functions and resources with the corresponding increases in the number of users.

 If we can engineer an application, which derives its system resources from its user-base, then otherwise economically infeasible services may be rendered feasible. The increase in the number of users would automatically scale up the resources of the application; allowing the service to support an increasing number of users without any degradation.

Ideally, the applications that may be set up in a Serverless mode would be the ones that allow real time interaction and sharing of information among the users. This selection is being made at this stage to solve the problem in a more pliable field. In the long run virtually any service may be delivered in a Serverless mode. Quoting from existing peer-to-peer applications, the IRC-chat and Naspster styled applications would be the easiest to implement in the Serverless mode since the servers in these cases are effectively limited to providing directory lookup services, brokering peer-to-peer connections, managing shared communication channels, and a range of peripheral services that assist with the real-time communications. Moving up the scale, hosting the web content in this mode would be more difficult since the servers in this case provide access to CPU and disk for (the CGI type) computations and content-storage respectively. There would be some server centric applications such as DBMS that will prove to be the hardest to implement, as Serverless applications, owing to their tight integration within the back-end server architecture and their performance requirements. The Serverless application architecture may thus be seen as the enabling technology that allows deployment of an Internet application without any specific need for a server.

The strategic advantage of this approach lies in the flexibility it would provide to the developers of such application where the only server based support they would require would be in the form of a download site. The users of the application would be able to download copies of the application and create a virtual application by pooling the resources of their individual computers. This approach would not only free the developers and the users of the application from finding support through advertisement revenues and/or other forms of user-pay based mechanisms but it will also open up access to a virtually unlimited system resource for hosting an application. Apart from simplifying the financial support model the Serverless application architecture would be a viable option in situations where only the basic network (IP-level) services remain available. In such a scenario the Serverless application would continue to operate when everything else has failed at the servers' level. The Serverless architecture also blends in with the mobile users requirements where a roaming device could find the appropriate services, through the Serverless applications installed on the mobile system, without requiring the local area networks to conform to its specific requirements. This form of use becomes even more relevant in the wireless networking environments where a mobile user may connect from anywhere and gain access to the desired service. The proposed application architecture would also remain highly resilient to the denial of service attacks.

An (Internet-wide) Serverless application may be defined as a service that exists solely by virtue of the instances of the application concurrently running, within the users' computers, anywhere on the Internet. Hence the challenges in implementing a Serverless application would lie in service discovery within the Internet, coordination of the distributed application system resources such as the CPU and the disk storage [1], the creation and management of a ubiquitous and distributed directory framework, an efficient sharing of information among the instances of the distributed application, the reliability of access, bandwidth conservation through the data-locality principle, the application security, and the application performance.

One of the key requirements, within any Serverless model, will be the ability to locate other active instances of the application within the Internet. In this regard we first need to look at the existing service location/discovery methods in order to evaluate their suitability for Serverless application architecture.

## 2. Contemporary Service Discovery Schemes

The current mode of operation of Internet is primarily unicast with relatively sparse pockets of multicast islands that exist in relative isolation. Initially, multicast was envisaged as a mean of streaming isochronous data, such as audio and video contents, within an enterprise [2]. Later, a wide-area multicasting architecture was proposed [3] which required the implementation of *mrouters* for carrying multicast packets over unicast networks. Multicast has the potential of being the key to the service discovery challenge however at this stage the penetration level of multicast is not sufficient and requires widespread adoption prior to being considered for use in an Internet-wide service discovery mechanism. Also, it has to be noted that multicast was primarily designed for sending data from a multimedia server to multiples destinations. Hence the inherent use is modeled for one-to-many data transmissions. In a true Serverless application architecture the pattern of communication would be on the basis of many-to-many. Hence the performance of multicast, for this relatively different mode of operation, would have to be evaluated to guard against multicast storms caused by a large number of intercommunicating Serverless application instances.

So far the work done in the service discovery has been limited to networks within an enterprise [4]-[7]. Presently, one of the widely known methods is the Service Location Protocol (SLP) by Veizade et al [4]. SLP too in its current form is limited to enterprise networks with shared services.

Hence making use of one these methodologies in a global context would require further work and a wide scale adoption among the Internet community. Also, it's worth pointing out that these methodologies are essentially server-based.

In the following sections a couple of Serverless models, addressing the service discovery and other related requirements, are presented for implementing the proposed application architecture.

## 3. Serverless Application Models

A higher-level generic Serverless application model must be established prior to the implementation of a possible architecture. We may opt for a model that solely exists for a new breed of applications developed specifically for Serverless application architecture. Alternatively, we may define a model that presents itself as a combination of the conventional servers and the virtual servers with the conventional application clients being able to access these servers transparently. A study of the former, type of the model, would assist in the development of the later. Hence in this paper the focus would be on the former, which will be referred to as Model 1.

Also, as mentioned earlier, the application architecture presented in this paper shall be limited to distributing the server-based functions, such as the directory lookups and connection management routines among the clients (i.e. the users' computers). The requirements of Model 1 shall be defined in the following sections and an application-architecture suitable for this model shall be presented.

### 3.1 *Service Discovery*

The life cycle of the Serverless application would commence with the first user downloading a copy of the application from the distribution web site. As part of the download process the web site would record the Internet address of the client and the domain name. The IP address is noted for discovering a client with a fixed IP address. However many of the clients would have dynamically assigned IP addresses hence the domain name information is also kept for performing a neighborhood search under the assumption that the client would reconnect from within the same domain. This assumption may not hold true in case the client changes the services provider or has multiple points of access (through different service providers). However this aspect would be dealt with at a later stage when we look at the benefits and the drawbacks arising from the increase in the number of the instances of the application.

After the download has been successfully completed, the web site adds this information to the application's directory entries component. The process of recording client information, within the application, is repeated every time a client downloads a copy of the application. The process is shown in Figure 1.

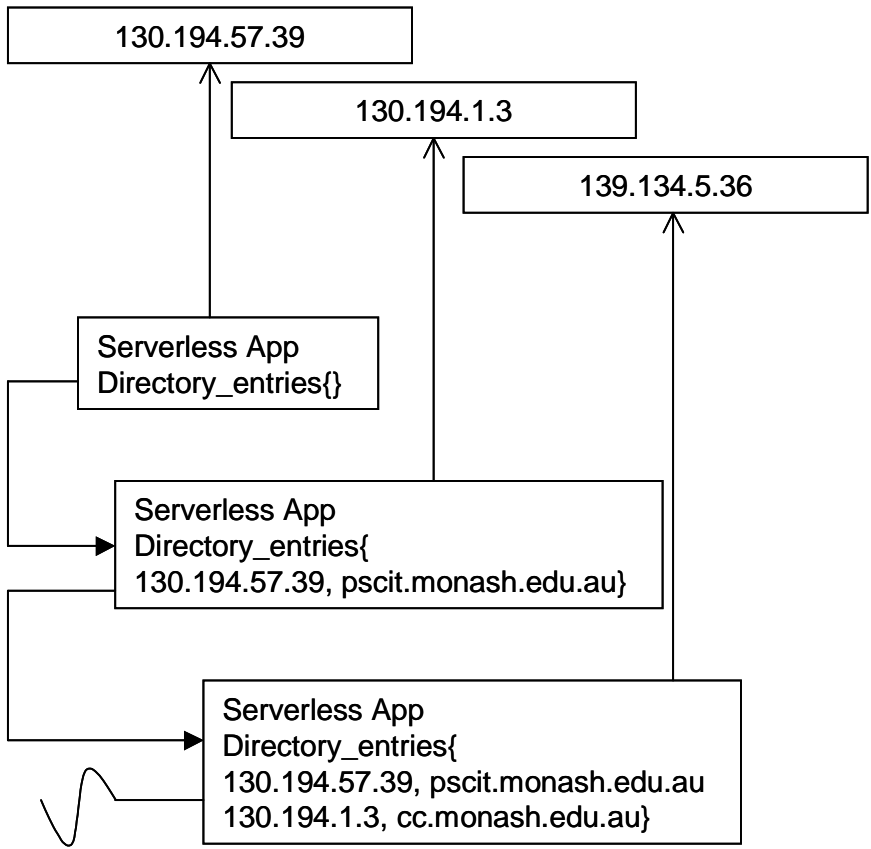

Figure 1: The client information encoding scheme.

### 3.1.1 Neighborhood Searching

The concept of neighborhood searching is taken from University of Arizona's Web Glimpse system [8]. One of the neighborhood concept defined in this web spider is based upon the link hops. The links in the document are traversed to a pre-defined limit (usually this is set to 2). This definition of the neighborhood is based upon the assumption that the links, within the web pages, are related to the page content. The neighborhood in the case of the downloading client would be the IP address range of the Internet Service Provider (ISP) in our case. Hence if the original IP address is no longer active then the clients may search the IP address range of the client's ISP. Doing so does not guarantee that the client, with an active instance of the application, would be found. However searching the IP range has a good chance of detecting the application if it's been configured for use and the client is active within the IP address range.

### 3.1.2  Scalability issues

The service location method, as described in section 3.1.1 would work as long as the predecessor clients remain within their respective IP address ranges and are active when a search is being performed. However with the increase in the number of clients the size of the directory entries would increase and scanning IP addresses would be prohibitively expensive in terms of time and bandwidth. Hence some changes to the application's directory entries scheme needs to be effected to limit the client searches. This may be achieved by satisfying the following conditions.

- Each instance of the application, running on a client, must have a map of the nearest applications based upon a sensible network metric. These maps are continually updated after being once populated.

- Newer instances of the application must identify themselves to their nearest neighbor with a populated map and obtain a copy of the same after adding its entry to the neighbor's map.

Implicit within the second condition is the flexibility that the directory entries made available to a downloading client may only contain a list of the nearer clients. A suitable number of entries may be fixed for this purpose. Thus we are now in a position to address the problem of a client having changed its Internet domain after downloading the application. For such a client one of the following conditions would apply:

- It is one of the earlier downloaded clients with very little or no information relating to its nearest neighbor being available.

- It's a client with a list of nearest neighbors that is no longer valid.

In the first case the client would need to wait to be polled by another client. In the second case the client must try to locate a router node for obtaining a listing of its currently active nearest neighbors. However in the first case the other clients may take a long time to discover the client or may not be able to discover the client at all if the client is in an isolated neighborhood. Hence it is important that a flagging mechanism is added to the neighborhood search mechanism where by such clients may be promptly detected. For this purpose the clients, including the router nodes, which wish to be contacted must advertise their particular in the designated web search engines. The active clients must, in addition to refreshing their maps, also consult with a specified list of web search engines to discover the identity of new clients. Hence the revised search scheme may be summarised in the following steps:

- Each instance of the application, running on a client, must have a map of the nearest applications based upon a sensible network metric. These maps are continually updated after being populated by referring to the router nodes. The router nodes must also periodically refer to the nominated search engines to check for the clients that might need to be included in the map.

- A new instance of the application must identify itself to one its nearest neighbor with a populated map and obtain a copy of the same. The instance should also list itself in the nominated search engines to facilitate discovery by the router node within the neighborhood. Once discovered and included into a neighborhood map, the instance should de-register itself from the search engines database provided it is not in the role of a router node. The router nodes must maintain a high visibility among the clients.

### 3.2  *The Serverless Application Attributes*

The distributed nature of the Serverless application requires that the information be exchanged among the instances of the application on regular basis. However not all information is required by all the instances hence different types of attributes need to be defined to cater for this requirement. In doing so the overall communication requirements may be optimised by adherence to the data locality principle. The following types of attributes would be initially required to manage the inter-application communication requirements.

- User defined attributes of an instance of the application, which are advertised within a specific group of instances.

- Local attributes of an instance of the application, which are advertised within the neighborhood (and are not visible to other neighborhood).

- Global attributes of an instance of the application, which are visible across all the neighborhoods.
- Common to the above would be the attribute update frequency. The application programmer would determine whether the attribute is refreshed/updated aggressively, moderately, or lightly. The actual rates corresponding to these categories would be dynamically computed on the basis of the prevalent network metrics within the neighborhoods.

Each instance of the application would be responsible for maintaining its own list of user defined, local, and global attributes. These attributes would be periodically exchanged within the target groups. The referential integrity of the attribute values would be ensured, within the target group, by only allowing an attribute value to be changed when all active instances in the target group have returned a receipt of the new attribute value. It is possible that an instance that is active at the time of the receipt of the update may go offline and hence fails to confirm its updated status; a time out mechanism would need to be effected to guard against such occurrences.

### 3.3  *Traversing a distributed map*

The nearest neighbor mapping would result in clusters of clients being created with the first hand information of the neighboring instances of the application. However in the Serverless application architecture any client may contact any other client anywhere in the Internet. Hence a communication scheme would need to be effected which allows any instance of the application within any neighborhood to be able to lookup and/or connect to any other instance based upon one or more shared attributes of the application. Hence the attribute values would have to be exchanged, periodically, within each neighborhood and also among the neighborhoods.

For example, an instance executing on a client within Neighborhood AA should be able to lookup and connect to another instance within Neighborhood BB on the basis of an application attribute shared between these two instances; a shared attribute may be a subscription to the same chat channel. This process is shown in Figure 2, where instance A of the application in neighborhood AA and instance B of the application in neighborhood BB share the same attribute value, i.e. `chat_group X`, with other instances of the application within neighborhood AA. This leads to the establishment

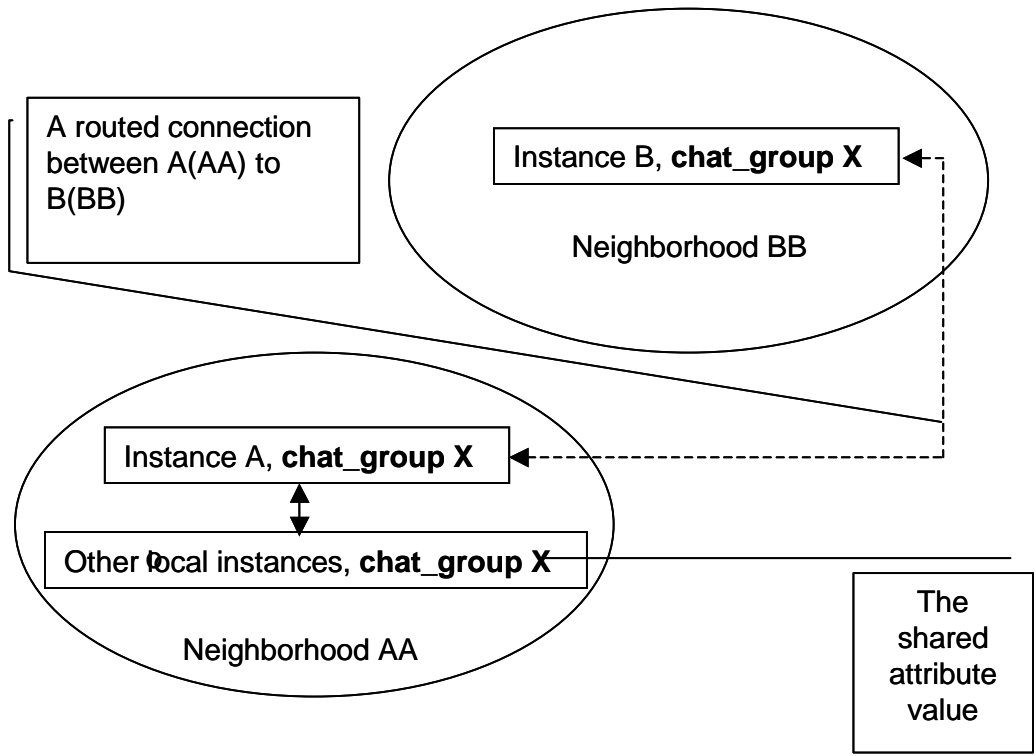

of a routed connection between instances A and B of the application.

Figure 2: Establishment of a routed connection between instances A and B.

A typical distributed (client) map would comprise the addresses of the active clients, their respective network metrics as measured from the neighborhood router, and the addresses of the router nodes in other neighborhoods. The router nodes and routed connections will be discussed in the following sections.

## 3.4 *Problems arising from the dynamic nature of the architecture*

It is quite possible that the clients assigned with specific group based tasks such as the clients acting as the router nodes may arbitrarily shutdown. This would require that the event, which would be the shutdown of the router node in this case, be detected and the next client in line to be able to take over the role of the router node. This arrangement may be effected through a beacon generation scheme where the qualifying clients actively compete, for the role of the router node, by monitoring a beacon signal.

## 3.5 *The routing requirements*

A large number of Serverless application instances running at arbitrary locations within the Internet, with varying levels of intercommunications needs, would require formulation of an efficient routing scheme that conserves the bandwidth by following the shorter paths and through active load-balancing. In the following section two of the main Internet routing protocols would be briefly reviewed to identify a suitable inter-application routing scheme.

## 4. Internet Routing Protocols

Internet is not static; connections may fail and later get replaced. Networks can get congested at one moment and then become underutlised the next. The purpose of route propagation mechanisms is not merely to find a set of routes, but to continually update the information [9].

### 4.1 *Distance Vector Routing*

The router keeps a list of all known routes in a table. At boot time, a router sets its routing table information to contain an entry for each directly connected network. Each entry in the table comprises a destination network address and the distance to that network, usually measured in hops. Periodically, each router sends a copy of its routing table to any other router it can reach directly. Gateway-to-Gateway [10] protocol was one of the earlier routing protocols based upon the distance-vector scheme.

### 4.2 *Link-State (SPF) Routing*

The distance-vector based routing methods require periodical updates comprising an entry for every possible network. The message size is proportional to the total number of networks on the internet being served by the router. Adding to this, the requirement of every router to participate in the update could enormously increase the volume of information thus exchanged.

One of the alternative to distance-vector algorithms is a class of algorithms known as *link state* or *Shortest Path First (SPF).* The SPF algorithm requires that each participating router to have complete topology information. The topology information may be graphically represented with each router corresponding to a node on a connected graph with the network that connect these routers as the edges of the graph. The nodes would share an edge if the two routers can directly communication with one another.

A router participating in an SPF algorithm performs two tasks. First, it continually updated the connectivity graph, also referred to as the map, by testing the links (the edges of the graph) connecting the neighboring nodes. Routers are neighbors if they share a link. Second, it propagates the link status information to all other routers.

To inform all other routers, each router periodically broadcasts a message that lists the status of each of its links. A status message, without specifying the routes, simply reports whether communication is possible between pairs of routers. On the basis of this information each router updates its topology information (map).

One of the main advantages of SPF algorithms is that each router computes routes independently using the same original data thus it is easy to achieve convergence in the route determining calculations. Finally, because the link status messages only carry information about the direct connections from a single router, hence the size does not depend on the number of networks in the internet. Thus, SPF algorithms scale better than distance-vector algorithms [11] [12].

## 5. Inter-application Routing

The Serverless application model embraces the concept of neighborhoods, which may be seen as equivalent to the groups of networks identified as the internets within the IP routing schemes. Hence, the key concepts within the routing protocols described in the previous section may also be utlised for managing the communications among instances of Serverless applications. In this regard the SPF algorithm with a complete map of the routing paths available to the routing node would be ideal to minimise the communication overhead. The formation of the routing maps could be made part of the service discovery scheme. Availability of a complete listing of Class A, B, and C networks addresses, such as the ones listed at [13], would help with the speeding up of the discovery and mapping processes. Alternatively, the outputs from `whois` [14] may be interpreted to determine the neighborhood of a client with a transient IP address. The maps thus generated may be advertised within the neighborhood. This process would also implicitly define the client's neighborhood membership. There are however some fundamental differences, between the proposed routing scheme and the IP routers, which need to be kept in mind:

The inter-application routing is to be handled by the running instances of the applications. Hence it will not be possible to have dedicated routing nodes within this environment. The most suitable nodes would perform the routing functions during their lifetime and then the tasks would be taken up by the next available set of nodes. Hence the assumptions valid for the dedicated platforms, such as the system resource remaining a constant entity, would no longer be so. Additional load balancing requirements would have to be taken into account for the variation in the capabilities of routing nodes.

Linked to the transient nature of the routing nodes will be the definition of an algorithm for replacing a routing node in the event of it becoming unavailable. This may be achieved by a local attribute that marks the continuing availability of the routing node. The node that would be the next in the list of forming a routing node would refrain from assuming this function as long as the attribute is there. In the absence of the attribute all nodes requiring router-mediated communications would hold back until an alternative attribute, from the replacement node, has become available.

It is also worth noting that not all communications among the instances of the applications, irrespective of their neighborhoods, would require assistance from the routing nodes. It is envisaged that the routing nodes would primarily act as the reference points for the new clients to find their way into the neighborhood, maintain the most up-to-date client map for the neighborhood, and to form the conduits for exchanging the attribute related information among the neighborhoods.

### 5.1 *Router node creation criteria*

One of the purposes of having the router nodes within a neighborhood would be to minimise the inter-application communication contents. However having a router node for a neighborhood with a handful of clients would not make any sense. In this case each of the client is better off managing its own communications. The router node may be created when the following criteria have been met.

- Hundreds of clients supporting active instances of the application are present in the neighborhood.
- There is a pool of clients that has a substantially high uptime. These would generally be the clients with permanent connections to the Internet.
- The network metrics are conducive for these clients to perform the router node functions. For instance a highly available node with a bottlenecked network connection may not be selected as a router node.

## 6. Stages in the life cycle of Serverless Application

As mentioned in an earlier section, the life cycle of the Serverless application would commence with the down load and creation of the first instance of the application. The following instances would discover their predecessors and form the first neighborhood. With an increase in the number of the instances, the neighborhood would be subdivided into smaller neighborhoods on the basis of data locality principle. The clients nearer to each other would be clustered into these newly formed neighborhoods. With a further increase in the number of the instances within the neighborhood, the routing nodes would be formed to assist with the communication processes. Thus providing better service times and lowering the risk of IP storms by preventing common nodes or clients from globally advertising their attributes and engaging into service discovery scans.

The only condition for a Serverless application to successfully deploy itself would be that there remains atleast one instance of application up and running, anywhere on the Internet, until a critical mass of instances has been formed. The critical mass would be the number of application that would require a subdivision of the initial neighborhood. Having the critical mass would ensure that there would always be a good chance of one or more instances running within the neighborhood. The numerical dimensions of a neighborhood shall be estimated in a later section.

The start up sequence of the application is shown in Figure 3. At time $T_1$ a client on the Internet download the first copy of the application and creates the first running instance of the application, $I_1$. Given the above assumption, this client must remain up for atleast as long as one more instance of the application has been made available within the neighborhood, at time $T_2$ another copy of the application is downloaded and the downloading client starts the second instance, $I_2$ of the application. $I_2$ has the information relating to the IP address and the domain of $I_1$. $I_2$ performs a search, as described in Section 3.1, and connects to $I_1$. $I_1$ add the IP address and the domain information of $I_2$ to its service location map. At time $T_3$ another instance of the application is created, which has the mapping information for $I_1$ and $I_2$. Owing to a lower network distance metric between $I_3$ and $I_2$, the former tries and connects to $I_2$ (and ignores $I_1$). The mapping information for $I_3$ gets conveyed to $I_2$ and $I_1$. By this time say $I_1$ has gone off the network hence the information needs be queued by the sender until the receiver re-instantiates the application or the internal time-out value is exceeded. Hence the onus of introducing itself, to the other instances, rests with the newer instance of the application. However the information is also be collected by the router node and through the search engines. At time $T_4$ instance $I_4$ gets created. This instance attempts to connect to $I_1$, owing to $I_1$ being its nearest neighbor. However, $I_1$ has logged off hence $I_4$ then tries to connect with $I_3$ and succeeds this time. This process is repeated for each new copy that get downloaded and installed.

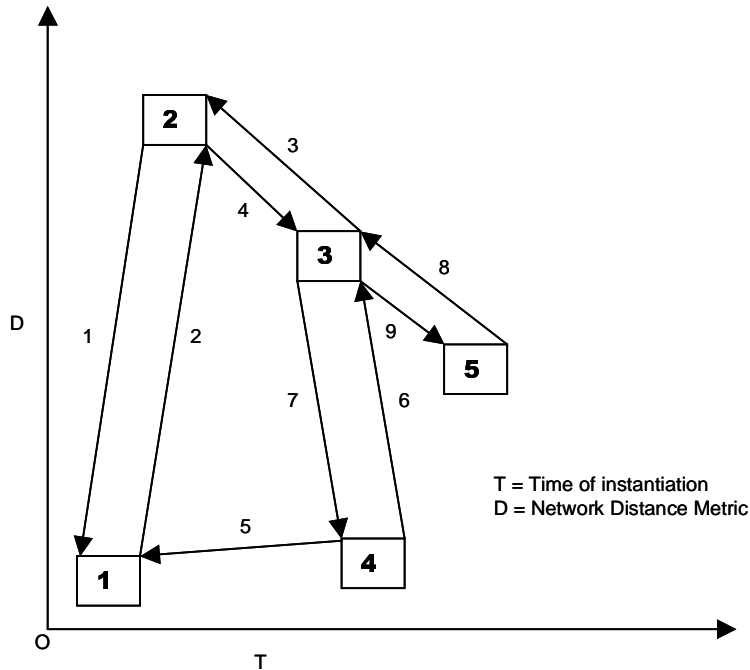

Figure 3: A start up sequence of Serverless application.

In the following section the method of exchanging information on regular basis among a large number of systems will be discussed.

## 7. Managing communications

Using one of the queuing techniques, as described in [15], we can find the distribution of waiting times for messages and the total time a message is in a network, waiting and being transmitted.

A basic form of queues is known as M/M/1 where the M stands for Markovian, after the mathematician Markov. The first letter M suggests that the messages arriving at the receiver end of the server is independent of one another; interarrival times are exponentially distributed. The second letter M suggests that the time to process the messages by the server, i.e. the service times, is also exponentially distributed. The number '1' indicates that there is a single server processing the queue.

### 7.1 *M/M/1 Queue*

The following assumptions apply in the case of M/M/1:

Message arrivals are independent of one another; the arrival times are exponentially distributed.

The service times are also exponentially distributed.

There is only one node (server) in the network.

If the link is busy when the message arrives at a node, it is placed in a queue until the link is free for onward transmission.

Using this method, the server can be either idle or busy, and the queue is expressed in terms of the number of messages waiting.

If G is the inter-arrival time then, the average rate of arriving messages can be expressed as:

$$A = \frac{1}{G}$$

If we have a message of length L and a link speed of B then the service time can be expressed as:

$$S = \frac{L}{B}$$

The Utilisation is expressed as a ratio of the arrival rate and the departure rate:

$$U = \frac{A}{D}$$

There are two possible states a network can be in using this method

There are more arrivals than departures

There are less arrivals than departures

Thus if the node is idle the utilisation is less than one, if we assume over a long period of time the number of messages arriving at the node will equal to the number of messages departing, then the probability that a node has a total number of $k$ messages in the system (queue + server) is:

$$P_k = (1-U)U^k$$

The number of messages in the system is also known as the state of the system.

Thus $P_k$ is geometrically distributed

$$P_k = kT_s$$

The waiting time is a function of the service time.

$$T_w = \frac{UT_s}{(1-U)}$$

From this we can see that for values of utilisation close to zero there is very little waiting time, and as the utilisation approaches 1, the waiting time becomes infinite.

The Total waiting time is.

$$T = T_s + T_w = \frac{T_s}{1-U}$$

The number of messages in the network

$$N = \frac{AT_s}{(1-U)} = \frac{U}{1-U}$$

The average number of messages

$$Q = AT_w = \frac{UAT_s}{(1-U)} = \frac{U^2}{1-U}$$

## 7.2  A global communication scheme

The attributes contained within any of the active instance of the application may vary at any time. For instance, in an Internet-chat styled instance the user may change the user identity or select a new channel to participate in. Hence the communication scheme should be able to propagate these changes among all the active instances of the application running anywhere within the Internet promptly. This may be achieved in an active mode or in a passive mode. In the active mode the instances, that undergo changes to their attributes, are allowed to commence the updating procedure at will. In the passive mode, the updating is only allowed to proceed, in an orderly fashion, at regular intervals of time.

The problem with the active mode may be highlighted by the following example. Assuming a relatively small sized neighborhood, comprising 256 active instances of the application, where each of the clients sends on the average of 64 bytes of data (equivalent of a default sized packet of ICMP ECHO_REPLY and excluding the IP header), to every other client within the neighborhood, per second. Even this seemingly minor rate of packet exchange has the potential to generate an average data rate of 128Kbits/sec at each of the client network interface. At present, most clients are equipped with 56Kbits/sec modem interfaces or less. Hence a more sensible means of exchanging information needs to be considered. The concept of partitioning the neighborhood would help however the critical mass for keeping the services running within each of the neighborhood may still require the presence of several hundred active clients. Hence these two conflicting requirements relating to the size of the neighborhood also need to be factored into the communication scheme.

Devising a communication scheme based upon divide and conquer principle may solve the problem of many-to-many communications occurring over the limited bandwidth interfaces. Under this scheme a neighborhood would be re-organised itself into smaller clusters of active clients. These clients may synchronise their attribute lists using a round robin scheme and elect a cluster leader, which will be the client with the lowest IP address value. Each cluster leader would then synchronise its attribute list with the other cluster leaders and re-distribute the final updated list among its cluster members. This process would then be repeated at regular intervals.

In order for this scheme to work all the clients in the neighborhood must have an up-to-date list of the clients within the neighborhood, to form non-overlapping cluster groups, and to be able to lookup the cluster leaders in adjacent clusters without having to exchange any messages.

The communication scheme for a neighborhood comprising 10 clients is represented in Figure 4. Assuming a cluster size of 5 in this case would force the neighborhood to be divided up into two clusters of five clients each. Each of the client, by referring to its listing of the neighborhood IP addresses, would be able to independently determine its cluster membership. Once the clusters have been formed the client with the lowest value of the IP address within the cluster would convey its attribute value to the next higher IP addressed client. The recipient client would add its own value of the attribute and convey it to the next higher IP addressed in the cluster. The client with the highest value of the address would add its own value to the list and return the finalized list to the sender. This process would be repeated until the lowest IP address in the cluster has received the copy of the finalized list. Once the attribute lists have been finalised at the cluster level, the cluster leader, with the lowest IP address, in the cluster leaders' group would propagate this information among other cluster leaders using a similar process followed within the clusters. The cluster leaders would then propagate the copy of the neighborhood wide finalised attribute list to other clients within their respective clusters. The last step where the cluster leaders propagate the neighborhood wide attribute list is not shown in the Figure for the sake of preserving clarity.

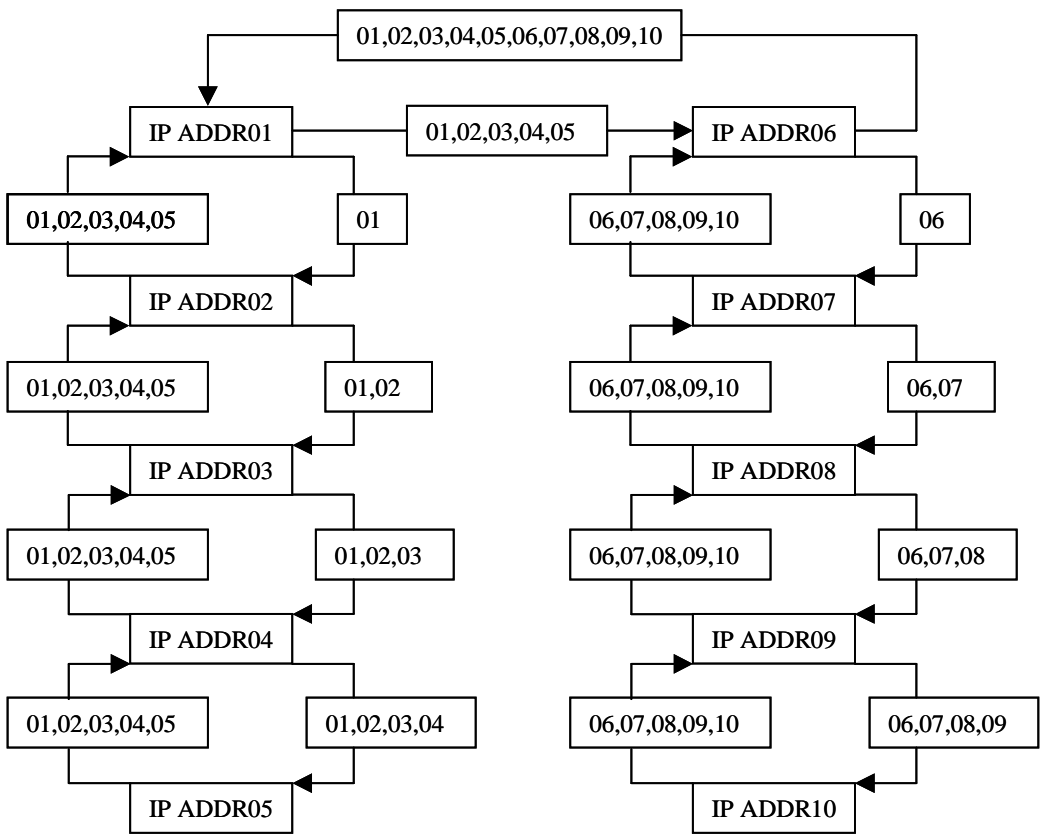

Figure 4: The hierarchical communication scheme for updating the attribute information among the instances of the application within the same neighborhood.

The proposed communication scheme allows concurrent updating of the information at cluster-level within the neighborhood. Hence the total time required to update the attribute list among all the clients within the neighborhood may be estimated by the following relationships.

$$T_u = t_c + t_{cl} + t_c^{'} \qquad (1)$$

$$t_c = \left| \sum_{i=1}^{2(n-1)} t_{ij} \quad j=1, N \right|_{max} \quad (2)$$

$$t_{cl} = \sum_{i=1}^{2(N-1)} t_i \quad (3)$$

$$t'_c = \left| \sum_{i=1}^{(n-1)} t_{ij} \quad j=1, N \right|_{max} \quad (4)$$

Where,

$N =$ is the total number of cluster leader nodes in the neighborhood.

$n =$ is the total number of nodes within each cluster (assuming all clusters have equal number of nodes).

$T_u =$ is the total estimated time for a neighborhood-wide update to be completed.

$t_c =$ is the maximum value of the times taken by the clusters in updating the information internally.

$t_{cl} =$ is the total time taken to update information among the cluster leader nodes.

$t'_c =$ is the maximum value of the times taken by the clusters in replicating the neighborhood-wide update, internally.

It may be noted that the proposed communication scheme differs from the normal use of M/M/1 queues in the respect that only a single message is propagated within the cluster for each update. Thus little of or no waiting time would be incurred, on this account, and the utilisation of the network bandwidth would remain close to zero. However the situation would change if more frequent updates were being performed leading to the service time for these requests to become greater than the waiting time.

The time taken to update the information within the clusters and for performing the neighborhood-wide updates would vary depending upon the state of the network links between the communicating nodes. Hence, to be able to ascertain the performance of a network we need to be able to estimate the load on the network, to do this we need to know the number and length of messages entering the network in order to predict the waiting and the service times; this may be hard to predict.

While keeping the above limitations in mind, the efficacy of the communication scheme may be determined using a statistical model. The governing relationships of the communication scheme, as given in equations (1) to (4), have been used for estimating the communication times. The timings obtained from the model are presented in the following section.

## 8. Communication Scheme Timings

The model is based upon the governing equations proposed in the pervious section. In this regard the clusters, within the neighborhood, are modeled as columns with the cluster leaders forming the top row. The model uses the concept of hops instead of the nodes considering there are $n-1$ hops for a cluster of member size $n$. Thus by keeping the array dimension to hops leads to simpler coding and of the model and a more compact representation. A message sent from one node to another node within the array is termed as a hop.

Following assumptions and approximations have been introduced to further simplify the model.

- A random time interval, ranging between 1-10, is assigned for each hop.

- No new random times are generated for the hops originating from the highest numbered IP address within the cluster. This is done based upon the assumption that the total random time intervals for the reverse propagation of the message within the cluster would generally be of an equivalent value. Hence the maximum of the total value for each cluster, while computing $t_c$, is multiplied with a factor of 2 to account for the propagation time required in sending the update from the highest addressed node in the cluster to the lowest addressed one. This leads to further code simplification.

- The pseudorandom generator does not use an exponential distribution. However this imperfection has been accepted on account of the proposed nature of the communication scheme where only a single message, at a time, gets propagated within the clusters and the cluster leaders.

The total communication time $T_u$ values, for globally updating information within neighborhoods comprising 256, 512, 1024, and 2048 hops, were calculated for varying cluster hop sizes (columns of the hops arrays) to determine the sensitivity of the scheme to varying rows and columns topologies. The results for the neighborhoods are tabulated in Tables 1-4 and Figures 5-8.

| Rows/Columns | $t_c$ | $t_{cl}$ | $t_c'$ | $T_u$ |
|---|---|---|---|---|
| 64/4 | 730 | 20 | 367 | 1118 |
| 32/8 | 399 | 40 | 202 | 641 |
| 16/16 | 223 | 82 | 107 | 412 |
| 8/32 | 125 | 143 | 60 | 328 |
| 4/64 | 66 | 369 | 31 | 466 |

Table 1: Communication timings for a 256 hops neighborhood.

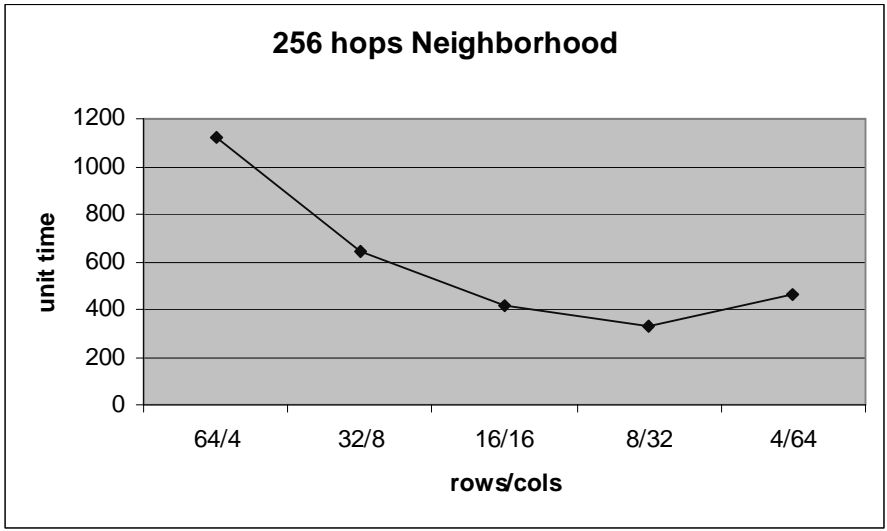

Figure 5: The total communication time $T_u$ for a 256 hops neighborhood.

| Rows/Columns | $t_c$ | $t_{cl}$ | $t_c'$ | $T_u$ |
|---|---|---|---|---|
| 128/4 | 1410 | 26 | 724 | 2160 |
| 64/8 | 789 | 34 | 373 | 1196 |
| 32/16 | 424 | 87 | 205 | 716 |
| 16/32 | 210 | 172 | 108 | 489 |
| 8/64 | 126 | 301 | 62 | 488 |
| 4/128 | 70 | 664 | 36 | 771 |

Table 2: Communication timings for a 512 hops neighborhood.

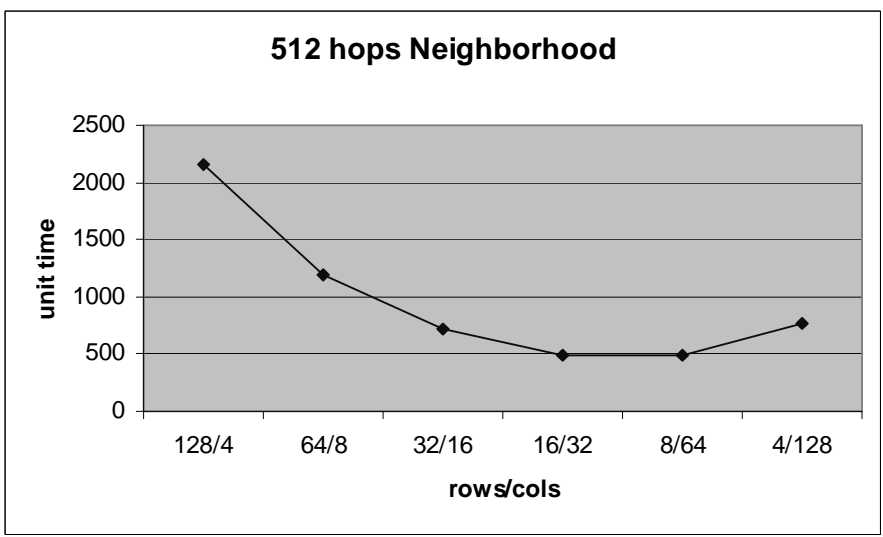

Figure 6: The total communication time $T_u$ for a 512 hops neighborhood.

| Rows/Columns | $t_c$ | $t_{cl}$ | $t_c'$ | $T_u$ |
|---|---|---|---|---|
| 256/4 | 2820 | 17 | 1438 | 4275 |
| 128/8 | 1437 | 53 | 786 | 2277 |
| 64/16 | 760 | 103 | 376 | 1239 |
| 32/32 | 413 | 195 | 205 | 813 |
| 16/64 | 220 | 359 | 113 | 692 |
| 8/128 | 130 | 700 | 61 | 891 |
| 4/256 | 71 | 1347 | 34 | 1452 |

Table 3: Communication timings for a 1024 hops neighborhood.

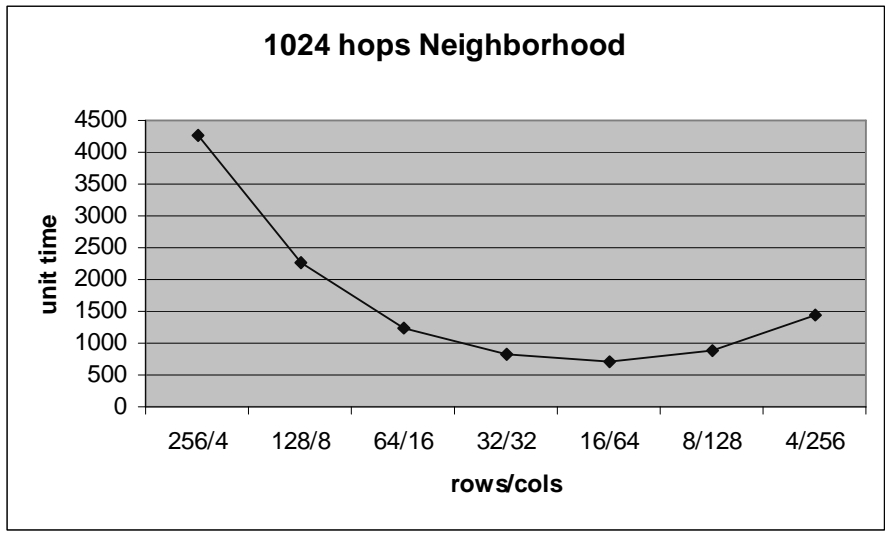

Figure 7: The total communication time $T_u$ for a 1024 hops neighborhood.

| Rows/Columns | $t_c$ | $t_{cl}$ | $t_c'$ | $T_u$ |
|---|---|---|---|---|
| 512/4 | 5705 | 26 | 2789 | 8520 |
| 256/8 | 2900 | 55 | 1464 | 4419 |
| 128/16 | 1547 | 95 | 748 | 2390 |
| 64/32 | 793 | 173 | 379 | 1345 |
| 32/64 | 408 | 345 | 221 | 975 |
| 16/128 | 230 | 759 | 114 | 1104 |
| 8/256 | 124 | 1448 | 62 | 1634 |
| 4/512 | 72 | 2801 | 36 | 2909 |

Table 4: Communication timings for a 2048 hops neighborhood.

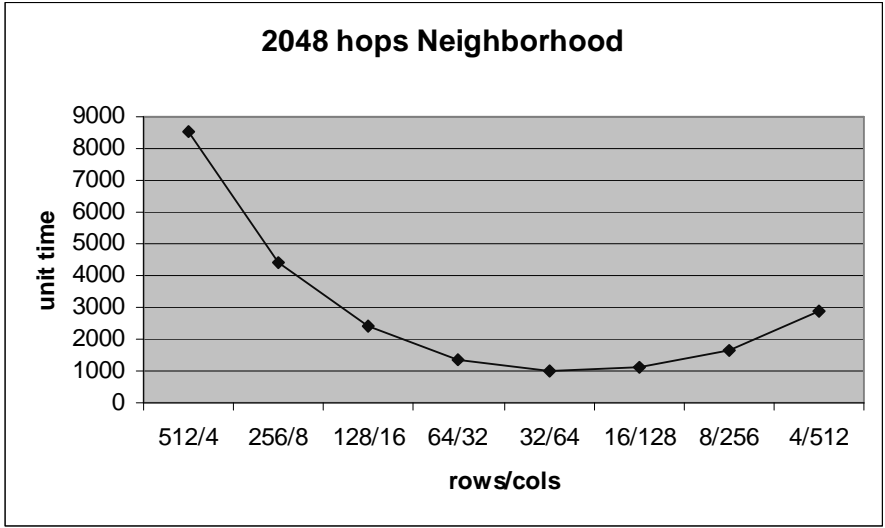

Figure 8: The total communication time $T_u$ for a 2048 hops neighborhood.

## 9. Comments and Recommendations

The total time for the proposed communication scheme, as shown in Figures 5-8, starts at a relatively higher value when the ratio of the number of the rows to the number of columns is the greatest or in physical terms shape of the array is rectangular; narrower at the base. The total time is found to be at the lowest values when the number of rows becomes less than the number of columns and the ratio is between the range of 0.25 to 0.5. The time then again starts to increase, as the shape of the array tends to become more rectangular, albeit wider at the base this time. However the rate of increase in the total time is considerably lower in the later case.

Hence based upon the results it may be predicted that the best communication topologies would be the ones with a relatively fewer nodes within the cluster and with a larger number of clusters in the neighborhood. Alternatively, this may be quantified by the ratio of clusters' hops to the cluster leaders' hops (rows and columns of the hops array):

$$0.25 \leq \frac{rows\_of\_the\_hops\_array}{columns\_of\_the\_hops\_array} \leq 0.5$$

The question regarding the optimal size of the neighborhood also needs to be answered in absolute terms. In order to do so, a workable time unit needs to be defined to replace the abstract unit times used in the experiment. In this regard the average time delay between the locally available systems and the regional/intercontinental systems needs to be estimated. For most cases this may be taken as follows.

$T_{dl} \leq 10ms$                                                                   **(5)**

$T_{dg} \leq 500ms$                                                               **(6)**

Where $T_{dl}$ is the time delay between two local systems and $T_{dg}$ is the time delay between regional systems, which may be estimated by noting the roundtrip timings for default sized ICMP ECHO_REQUEST/ECHO_REPLY packets.

The unit time for sending a message between any two nodes, within the computational model, is allowed to randomly vary between a range of 1 to 10 time units. Hence mapping the above estimated Internet time delays to the time range of the model from the higher valued expression equates one unit time to 50 ms of Internet time delay.

Hence the minimum values derived from the simulations may now be assigned real-time values. These values are shown in Table 5.

| Hops Array Size | Optimal Dimensions | $T_u$ (Milliseconds) |
|---|---|---|
| 256 | 8 x 32 | 16,400 |
| 512 | 8 x 64 | 24,400 |
| 1048 | 16 x 64 | 34,600 |
| 2048 | 32 x 64 | 48,750 |

Table 5: Total communication times for the arrays in milliseconds.

The results show that the communication scheme would globally update information within neighborhoods comprising 256, 512, 1048, and 2048 hops in 16.4, 24.4, 34.6, and 48.7 seconds respectively. The relationship between the number of hops within a neighborhood and the time to propagate a neighborhood-wide update is shown in Figure 9.

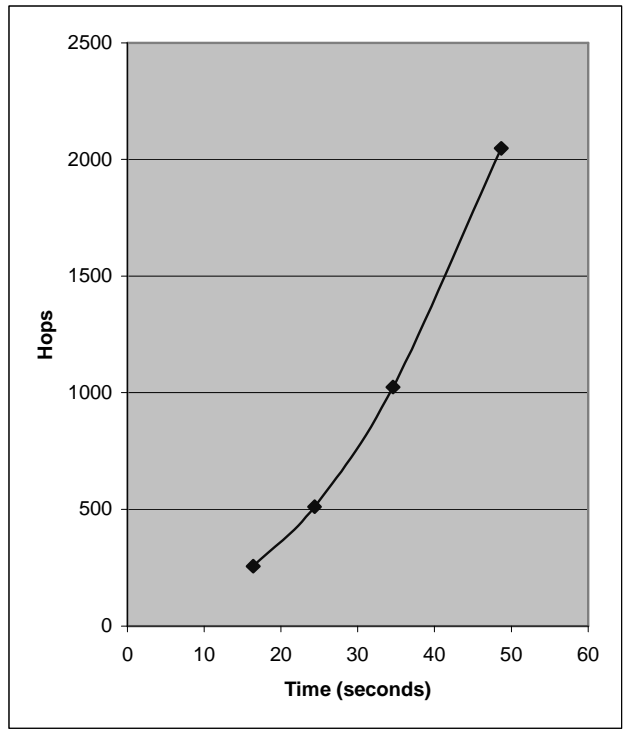

Figure 9: The relationship between the neighborhood-wide update time and the number of hops within the neighborhood.

It may be noted that the mapping of 50 milliseconds per unit time corresponds to the worst-case Internet time delay as estimated in expressions (5) and (6).

It is worth conceding at this point that the global communication scheme presented in the paper is a rather basic one developed for a Transputer array [16]. Hence there is a considerable room for improvement. Also, based upon the scalability exhibited within this scheme, it would be reasonable to assume that much larger sized neighborhoods may be serviced efficiently through the proposed application architecture by developing more sophisticated communication schemes.

## 10. Concluding Remarks

The proposed models embrace an infrastructure that is inherently random in nature owing to its reliance on a multitude of autonomous computing systems linked through varying types of networks; hence there will be times when some of the information may be not be readily available or it may not become available in the manner we would normally expect it to be. In other words we will need to get used to the idea of the applications behaving in a semi-deterministic manner (unlike the conventional ones with fully deterministic responses). The closest example would be that of Napster where access to a particular computer file is not guaranteed, but users generally tend to get what they want. The proposed model merely keeps the uncertainty levels within a reasonable limit and ensures that the level of uncertainty reduces with an increase in the number of unreliable systems (rather than increasing as a consequence of it). There is no concerted effort made to entirely eliminate the element of uncertainty from the model.

The proposed communication scheme is demonstrated to be viable for deploying the Serverless applications. However the contents of the intercommunications between clients running the instances of the application, the updates among the neighborhoods, and the communication scheme for the routers still needs to be investigated. It is also desirable that the proposed communication scheme be tested in a real environment to verify the results predicted by the model.

The focus of research in this paper is primarily at the start-up phase of the applications where relatively fewer and sparsely located clients would need to interconnect and synchronise. However as the number of active clients reaches in the order of millions then some of the redundant measures proposed for ensuring service discovery and map updates may no longer be necessary. However by the same token the requirement of optimising inter-application communication shall become more critical.

The paper by no means claims to solve all the challenges of the proposed architecture. At this stage it merely attempts to chart a path for future research and development for this architecture and presents an over view of the challenges ahead in the development of applications based upon this architecture. The main areas that need to be investigated are summarised to assist in directing future research and development efforts:

- Serverless applications models. A couple of models have been proposed in the paper however more refined models would have to be developed to meet the diverse requirements of future networked applications.

- Service discovery within the Internet, particularly relating to dynamically assigned

and transient IP addresses, is a vast area and the coverage in the paper is by no means an exhaustive one. More work would be needed for the development of efficient (Internet-wide) Serverless search protocols.

- The protocols for communicating attributes among the instances of Serverless applications would need to be defined by identifying the critical information fields required by different types of applications. For instance the attribute information that needs to be exchanged for Internet-chat type of application may be different from the one that provides a content-based service (such a distributed file storage space).

- The methods of jump-starting the Serverless applications need to be improved upon. One of the weaknesses in Model 1 is that initially there may not be enough active instances of the application to attract the users. Hence further work must include Model 2 styled architecture where it is possible to integrate the existing server-based applications within the Serverless architecture.

- The testing of the proposed communication model needs to be done using active clients to verify the results.

- The inter-neighborhood communications have to be modeled and tested to complete the communication model for the entire Serverless application domain (i.e. the collection of the neighborhoods).

- Distributed file storage made available by the participating Serverless clients is an area that has not been discussed in the paper. Providing file storage in this manner has the potential of widening the scope of the Serverless application architecture considerably.

- Finally, the use of agent-based technologies may be investigated for managing the inter-application operational requirements.

## Acknowledgements

The authors would like to thank Patrik Mihailescu for refining some of the key concepts presented in the paper.